\documentclass[amsmath,amssymb,superscriptaddress,notitlepage,twocolumn,nofootinbib]{revtex4-1}
\pdfoutput=1

\usepackage{graphicx}
\usepackage{dcolumn}
\usepackage{bm}

\newcommand{\mnras}{{\em Mon. Not. Roy. Astron. Soc. }}

\newcommand{\physrep}{{\em Phys. Rept. }}

\newcommand{\aj}{AJ }

\begin{document}

\title{Dipole of the Epoch of Reionization 21-cm signal}

\author{An\v{z}e Slosar}
\email{anze@bnl.gov}
\affiliation{Brookhaven National Laboratory, Upton, NY 11973, USA}

\date{\today}

\begin{abstract}
  The motion of the solar system with respect to the cosmic rest frame
  modulates the monopole of the Epoch of Reionization 21-cm signal
  into a dipole. This dipole has a characteristic frequency dependence
  that is dominated by the frequency derivative of the monopole
  signal. We argue that although the signal is weaker by a factor of
  $\sim100$, there are significant benefits in measuring the dipole.
  Most importantly, the direction of the cosmic velocity vector is
  known exquisitely well from the cosmic microwave background and is not
  aligned with the galaxy velocity vector that modulates the
  foreground monopole. Moreover, an experiment designed to measure a
  dipole can rely on differencing patches of the sky rather than
  making an absolute signal measurement, which helps with some
  systematic effects.
\end{abstract}

\maketitle

\section{Introduction}

The earliest direct image of the universe comes from the observations
of the Comic Microwave Background (CMB), which arises when photons and
decouple from cosmic plasma a few hundred thousand years after the big
bang at a redshift of $z\sim 1100$. The universe then enters a period
known as ``dark ages'', where neutral hydrogen slowly cools and
collapses into halos, but the first stars have not yet ignited. The
first luminous object form at redshifts of $z\sim 20-40$, but very
little is actually known about this early period. With time, galaxies
form and start filling the universe with photo-ionizing radiation
which re-ionizes the hydrogen in the inter-galactic medium in the
process that is thought to have completed by redshift of around
$z\sim 6$. This period in the evolution of the universe is known as
the epoch of reionization (EoR).  It is thought that structure in the
universe during this period is characterized by growing bubbles of
ionizied hydrogen surrounded by yet-to-be-ionized neutral hydrogen. The
neutral hydrogen shines in radio in the 21-cm hydrogen
line. Measurements of the redshifted 21-cm line are thus thought to be
the most promising way of constraining reionization (EoR)
\cite{2006PhR...433..181F,Morales2009:0910.3010v1,Pritchard2011:1109.6012v2}. They
will teach us both about the astrophysics of this complex era in the
evolution of the universe, as well as provide strong constraints on
the value of the total optical depth to the surface of
last-scattering, which will help with measurements of many
cosmologically relevant parameters, most importantly the neutrino
mass\cite{Liu2015:1509.08463v2}.

Up to now, most experiments in the field have focused on either
measuring the fluctuations in the 21-cm line by measuring the 21-cm
brightness temperature and relying on the foreground smoothness to
isolate it
\cite{Tingay2012:1212.1327v1,Parsons2009:0904.2334v2,Rottgering2006:astro-ph/0610596v2,Paciga2013:1301.5906v2},
or on attempting to measure the global signal, the monopole of the
21-cm radiation from the EoR
\cite{Monsalve2016:1602.08065v1,Ellingson2012:1204.4816v3,Greenhill2012:1201.1700v1}. The
latter measurement is tempting, since the signal is relatively strong
and simple back-of-the envelope calculations show that it should be
easily achiveable based on SNR calculations. However, the experimental
challenges are daunting: the foregrounds are brighter  than the signal by orders of magnitude 
and vary very strongly across the sky, which makes calibration of the
instrument and beams to the required level of precision is very
difficult.

In this note we make a very simple point that one could attempt to
measure the dipole of the signal rather than the monopole. Although
the signal is reduced by a factor of around 100, the systematic gains
are very significant. The problem is in many ways analogous to the
Cosmic Microwave Background -- measuring Cosmic Microwave Background
(CMB) dipole is significantly easier than measuring the CMB monopole
or the CMB temperature fluctuations.  However, one should not take
this analogy too far for two reasons.  First, while the sky signal on
large scales is dominated by the CMB at frequencies above $\sim$ 1GHz,
this is not true for the 21-cm EoR signal: a total dipole is going to
be dominated by the foregrounds by order of magnitude at the relevant
frequencies. Second, while the dipole signal in the CMB is two orders
of magnitude larger than the higher order mulitpoles, the same is not
true for the the EoR signal, which has comparable or higher power at
degree scales compared to dipole. Nevertheless, as we will discuss in
this paper, the dipole measurement still has several attractive
features in regards of systematic effects.

\section{The signal}

While the details are poorly know, the general outline of the process
of reionization and the general features of the evolution of the 21-cm
brightness temperature with cosmic time are understood.   We will not go
into detail, but refer reader to well know reviews
\cite{Pritchard2011:1109.6012v2}. 

\begin{figure*}
  \centering
\includegraphics[width=\linewidth]{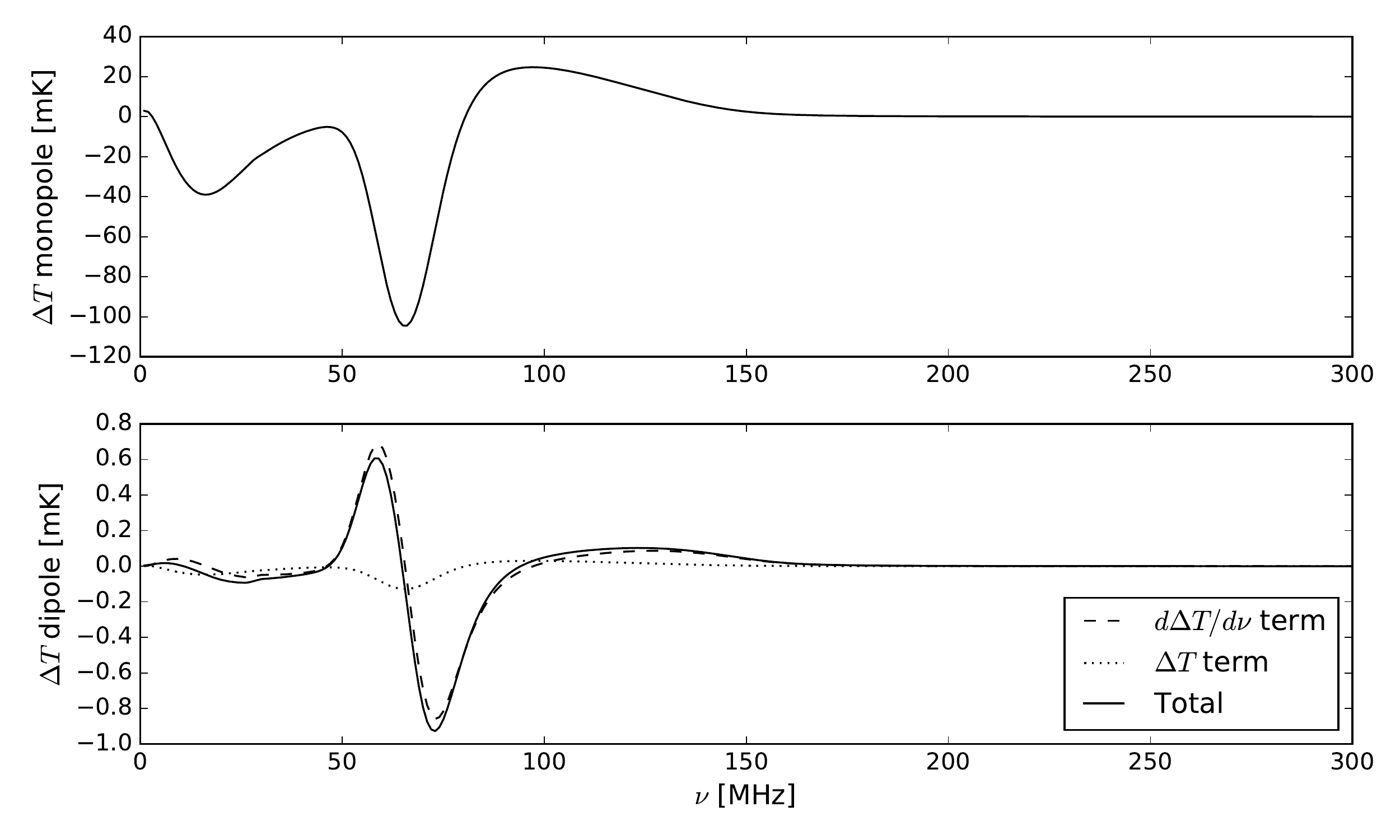}
\caption{The figure showing the the EoR monopole and dipole. The
  monopole follows the model of \cite{Pritchard2010:1005.4057v1}. Dipole
  has two terms, one proportional to the monopole and the other to its
derivative in frequency direction. The latter dominates the signal. \label{fig:1}}

\end{figure*}

The upper plot of the Figure \ref{fig:1} shows the 21-cm global signal
for a popular model. The monopole of the EoR signal is always observed
relative to the CMB monopole and is sometimes seen in absorption and
sometimes in emission. The magnitude of the observed signal is
determined by the difference between the CMB and spin temperatures at
a given redshift, the latter being the excitation temperature given by
the relative occupancy of the two 21-cm states. Depending on the
epoch, the gas is observed sometimes in absorption and sometimes in
emission. The spin temperature is determined by absorption/emission of
CMB photons, collisions with other species and resonant scattering of
the Lyman-$\alpha$ photons. At very high redshifts $z \gtrsim 200$,
the spin temperature is still thermally coupled to CMB via residual
Compton scattering and therefore the expected signal is zero. When
this process becomes inefficient, the spin temperature becomes
collisionally coupled to gas, which cools adiabatically as
$\propto (1+z)^{-2}$ and so is seen in absorption compared to CMB that
cools $\propto (1+z)^{-1}$ (the first through in the upper panel of
Figure \ref{fig:1}). At redshifts $z \sim 40$, gas becomes to
rarefied for collisional coupling and radiative coupling brings spin
temperature back to radiation temperature, erasing the signal. When
first sources appear at $z\sim 20$, they emit Lyman-$\alpha$ and X-ray
photons, which re-couple spin temperature to gas temperature via
Wouthuysen–Field effect \cite{1952AJ.....57R..31W,1958PIRE...46..240F}. However,  at that epoch, the gas
is still colder than CMB resulting in a second bout of 21-cm being
observed in absorption (the second through in the upper panel of
Figure \ref{fig:1}). Later, Lyman-$\alpha$ coupling saturates and
the gas temperature rises above radiation temperature, giving rise to
overall signal in emission. At this complex period, there are large
variations in the signal across space and the total emission is driven
by fluctuations in ionization, density and gas
temperature. Eventually, the universe reionizes and the mean signal
drops back to zero because majority of intergalactic gas is
ionized. At even lower redshifts, 21-cm is detected in pockets of
neutral hydrogen in galaxies.

Motion of the Earth with respect to the cosmic rest frame modulates
the monopole of the signal via two separate effects: i) the frequency
independent boosting in source intensity by $v/c$ factor (i.e. the effect
that generates a temperature dipole from monopole in CMB), ii) the
blueshifting of photons in frequency by  $1+v/c$ factor (i.e. moving
towards CMB, at fixed frequency we're observing photons from
lower-frequencies blue-shifted into our band). 
For clarity, let's write the monopole signal as a sum of frequency
independent and frequency dependent parts $T_{\rm mono}(\nu) =
T_0+\Delta T(\nu)$, where the frequency independent part contains all the
large frequency fixed signals we know exist (e.g. CMB).   The total observed
signal from Doppler shifting of the monopole is thus given by

\begin{multline}
  T_{\rm obs}(\theta,\nu) = \left(T_0 + \Delta T (\nu-\delta\nu)\right)
  \left(1+\delta \nu/\nu) \right) = \\
  T_0+\Delta T (\nu) + \left( T_0 + \Delta T(\nu)- \frac{d \Delta T}{d \nu} \nu \right) \frac{v_d}{c}\cos \theta,
\end{multline}
where $\delta\nu /\nu=v_d/c \sim 1.2\times 10^{-3} $ is
the amplitude of the velocity dipole and $\theta$ is the angle with respect to the
velocity vector of our motion with respect to the cosmic rest frame.
The dipole signal is thus given by
\begin{equation}
  T_{\rm dip}=\left( T_0 + \Delta T(\nu)- \frac{d \Delta T}{d \nu} \nu \right) \frac{v_d}{c}\cos \theta.
\end{equation}

We see that the dipole signal has three components (corresponding to
three terms in brackets above): the traditional frequency independent
dipole, which would match the CMB dipole in the absence of
foregrounds, the traditional boosting of the frequency independent
signal due to Doppler shift and also a term that takes into account
the frequency dependence of the EoR monopole signal. We plot both
frequency dependent contributions in the Figure \ref{fig:1}. We see
that the derivative signal dominates the signal. The total signal has
the amplitude of about $0.5$mK.

\begin{figure*}
  \centering
\begin{tabular}{ccc}
\includegraphics[width=0.33\linewidth]{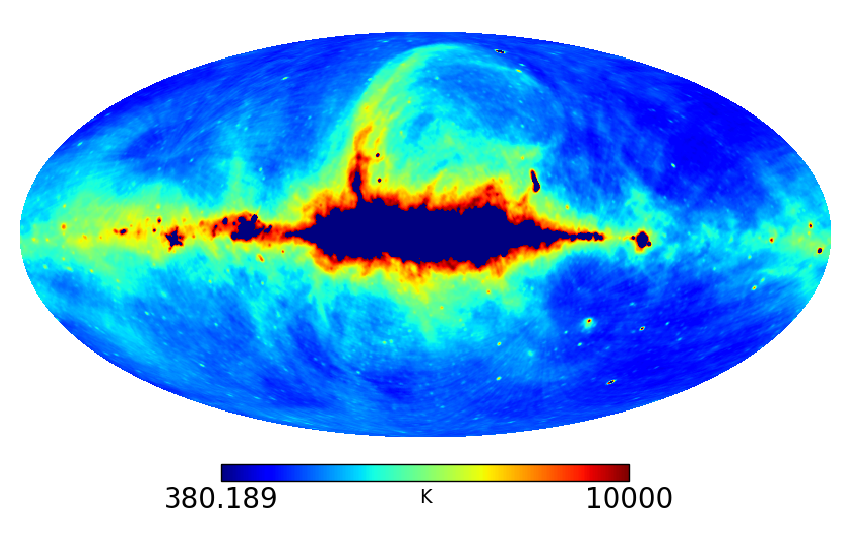} &
\includegraphics[width=0.33\linewidth]{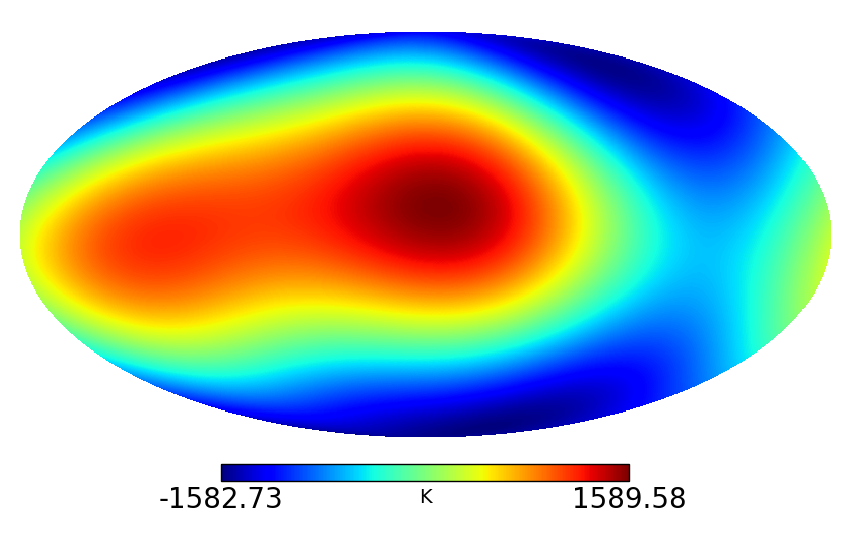} &
\includegraphics[width=0.33\linewidth]{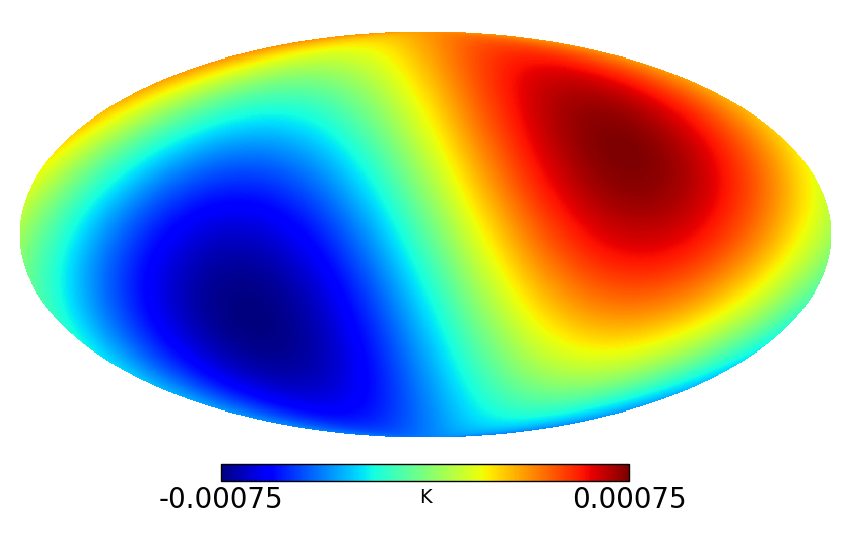} \\
\end{tabular}
\caption{Estimated map of foregrounds at 60MHz using Global Sky Map
  \cite{2008MNRAS.388..247D} masked where it exceeds 10,000K (left), dipole and quadrupole of the
  above map (which are likely be confused for a finite sky covarage
  experiment (middle) and the map of the expected signal (right). \label{fig:2}}
\end{figure*}

\section{The foreground question}
The foregrounds, of course, are what is really difficult about these
measurements. To give an impression of how difficult these can be, we
plot a rough estimate of the foreground on Figure \ref{fig:2} at
60MHz. This figure is based on the Global Sky Models (GSM) from
\cite{2008MNRAS.388..247D}. We have masked pixels with temperature
above $10^4$K, since an experiment with a finite angular resolution
would be able to optimally downweight bright parts of the sky. The
middle panel shows the dipole and quadrupole of the foreground while
the right panel shows the estimated signal at the same frequency. We
note that the constituent datasets that enter the GSM have uncertain
fidelity on the largest scales and therefore the plotted foreground
monopole and quadrupole might be significantly off.

Note that one such map exist at every frequency. While a search for
monopole can marginalise over the foreground model while subtracting
it\footnote{In general, monopole experiments also rely on spectral
  smoothness of the foregrounds, but note that subtracting and
  marginalising foreground model subject to spectral smoothness prior
  is conceptually the same.}, in dipole, the same marignalisation can be done subject to
constraint that the resulting dipole is aligned with the cosmic dipole
at every frequency. This is a very informative prior.

Methods for self-calibrated foreground rejection that rely on the
fact that monopole does not vary across the sky, while the foregrounds
do\cite{2014ApJ...793..102S}, can be easily generalised to dipole,
since the dipole varies across the sky in a precisely known fashion.

\section{What experiment would measure this}

The most promising design the measuring this signal would be a
differencing radiometer measuring the difference between two widely
separated points on the sky. Unfortunately, unlike the CMB, where the
signal above $\sim$ 1GHz is domimated by the CMB monopole, the
foregrounds will dominate for EoR dipole measurement. This calls for a
sufficient angular resolution to resolve the radio loud and radio
quiet parts of the sky in order to allow optimal weighting
\cite{Liu2012:1211.3743v3} and foreground rejection \cite{2014ApJ...793..102S}.
Moreover, radio-loud parts of the foreground sky can be used to
characterise the frequency response of the receiver antenna.

The usual techniques used in CMB instrumentation could be used to
inoculate against most common systematic: by putting the two receivers
on a platform that rotates sufficiently fast, one can calibrate the
beam differences between the two horns and by using 180$^\circ$
hybridisation one can remove the receiver $1/f$ noise.  Note that
while variations of these techniques can be used in the monopole
measurement, they are considerably less efficient. Since the noise
will always be dominated by the sky noise there is no need for
cryogenicaly cooled receivers.

What sensitivity would be required to perform this measurement?  A
convincing and accurate forecast would need to start with a mocked up radio-sky,
including signal and realistic foregrounds, simulate observed maps
with a realistic window function and then apply inverse covariance
weighting to optimally extract the signal. This clearly exceeds the
scope of this paper. Instead, we will make a back-of-the-envelope
calculation to demonstrate that noise properties of a reasonable
experiments can achieve desired statistical sensitivity. 

We start with a radiometer equation that tells us that the error on
measurement of the noise temperature is given by 
\begin{equation}
  \Delta T = \frac{T_{\rm sys}}{\sqrt{\Delta \nu t}},
\end{equation}
where $T_{\rm sys}$ is the system temperature, $\Delta \nu$ is the
observing bandwidth and $t$ is time to observe.  Somewhat
counter-intuitively, the receiving area does not come into this
equation, since for a uniform unresolved radiation, the bigger
collecting area is exactly canceled by a smaller beam-size. In our case we
do want sufficient resolving power to be able to isolate radio-quiet
and radio-loud parts of the foregrounds.

The radio sky at $60$MHz varies between $2,000$K and $40,000$K. From
statistical perspective, one would just choose two quietest patches of
the sky, however a scanning experiment does not have much freedom in
choosing which parts of the sky to observe and besides more sky leads
to better systematics control. But because we can still downweight
radio-loud parts of the sky, it is not optimistic to assume just a
uniform sky temperature of $10,000$K. At this level, the noise
properties of receivers are irrelevant.

An experiment would measure the signal in many small frequency bins
and the total signal to noise is given by an integral of observed bandwidth
\begin{equation}
  {\rm SNR}^2 \sim \frac{N_d t}{2} \int_{\nu_{\rm min}}^{\nu_{\rm
      max}} \left(\frac{\Delta T_{\rm dip}(\nu)}{T_{\rm sys}(\nu)}\right)^2 d\nu,
\end{equation}
where $N_d$ is the number of receiving elements and the factor of 2
accounts for the fact that amplitude of the dipole accounts for
maximum temperature difference, not typical one. Note that there could
be extra factors of two, depending on the exact differencing
scheme.  Assuming an experiment operating between 50MHz and 100MHz, we
find that 15 element radiometer could measure the signal at about
$5\sigma$ over a course of a year. This result of course crucially
(quadratically) depends on the assumed $T_{\rm sys}$. Assuming that
weighting data optimally can bring the effective temperature to
$5000$K, only four elements would suffice\cite{Liu2012:1211.3743v3}.

\section{Discussion \& Conclusions}

Differencing has proven to be one of the most successful paradigms in
the experimental physics: differential measurements are easy, absolute
measurements are hard. We apply this principle to the problem of
measuring the EoR monopole. Due to our motion with respect to the
cosmic rest frame, this signal is modulated in a dipole fashion.  The
amplitude of this dipole is supressed but somewhat less than $v_d/c$
factor due to a non-trivial frequency structure of the signal. This
supression of the signal could be more than compensated by
considerably easier systematic control in the dipole measurement:
\begin{itemize}
\item The direction of the CMB dipole is know very well and more
  importantly, the galactic foreground will have both  a different
  true dipole and the doppler dipole of the foregrounds will be
  different: motion of the solar system with respect to the CMB is not
  the same as its motion with respect to the galaxy. This can be used
  to estimate the residual foreground contamination.

\item The standard differencing techniques well known in the radio
  astronomy can be used to great advantage in this set up. This should
  help in dealing with radio frequency interference, the amplifier
  $1/f$ noise and the earth's atmosphere. However, the mean beam
  chromaticity will remain a significant issue.

\item The signal derived in this way could be used to cross-check
  measurements derived from the monopole, since the information
  content is the same. In fact, one could imagine an experiment that
  would measure both at the same time. 

\item Since the signal is proportional to the derivative of the
  monopole with respect to the frequency, this technique could be
  potentially very efficient for reionization scenarios that happen rapidly.
\end{itemize}

We have made a back-of-the-envelope estimate of the require
signal-to-noise and determined that signal is in principle measurable
in a reasonable amount of time for a reasonable experiment. We hope
that this warrants a more accurate forecasts, which would take into
account the spactial and frequency variation of foregrounds and work out
an optimal map-making scheme.

\appendix
\section*{Acknowledgements}

I thank Adrian Liu for providing numbers that were used to make Figure
\ref{fig:1}. I acknowledge useful discussions with Uro\v{s} Seljak,
Ue-Li Pen, Eric Switzer, Chris Sheehy and Paul Stankus.

\bibliography{cosmo,cosmo_preprints}

\end{document}